\documentclass[galaxies,article,accept,pdftex,moreauthors]{Definitions/mdpi} 
\firstpage{1} 
\pubvolume{1}
\issuenum{1}
\articlenumber{0}
\pubyear{2023}
\copyrightyear{2023}
\externaleditor{Academic Editor: Margo Aller} 

\datereceived{{24 January 2023}} 
\dateaccepted{{4 March 2023}} 
\datepublished{{}} 
\hreflink{ https://doi.org/10.3390/galaxies11020041 } % If needed use \linebreak
\pdfoutput=1

\usepackage{pdflscape}	% Landscape pages
\usepackage{mathtools}
\usepackage{amsmath}
\usepackage{adjustbox}
\usepackage{multirow}
\usepackage{upgreek}
\usepackage{amssymb}
\usepackage{graphicx}
\usepackage{graphics}
\usepackage{multirow}
\usepackage{color}
\usepackage{xcolor}
\usepackage{bm}	
\usepackage{comment}
\usepackage{upgreek}
\usepackage{float}
\usepackage{graphicx}
\usepackage{amsmath}
\usepackage{adjustbox}
\usepackage{multirow}
\usepackage{upgreek}
\usepackage{graphicx}
\usepackage{graphics}
\usepackage{multirow}
\usepackage{color}
\usepackage{xcolor}
\usepackage{bm}	
\usepackage{comment}
\usepackage{upgreek}
\usepackage{float}

{\newif\ifnotend
\notendtrue
\def\veclist{ABCDEFGHIJKLMNOPQRSTUVWXYZabcdefghijklmnopqrstuvwxyz.}
\def\top#1#2.{#1}
\def\tail#1#2.{#2.}
\loop\expandafter\xdef\csname v\expandafter\top\veclist\endcsname%
{{\noexpand\bf\expandafter\top\veclist}}
\edef\veclist{\expandafter\tail\veclist}
\if\veclist.\notendfalse\fi\ifnotend\repeat}

\def\km2s2{{\rm\,km^2\,s^{-2}}}

\def\ltsima{$\; \buildrel < \over \sim \;$}
\def\simlt{\lower.5ex\hbox{\ltsima}}
\def\gtsima{$\; \buildrel > \over \sim \;$}
\def\simgt{\lower.5ex\hbox{\gtsima}}

\Title{Gravitational Instability of Gas--Dust Circumnuclear Disks in Galaxies}

\TitleCitation{Gravitational Instability of Gas--Dust Circumnuclear Disks in Galaxies}

\Author{Roman 
 Tkachenko $^{1,}$*$^{,\dagger}$\orcidA{}, Vladimir Korchagin $^{1,\dagger}$\orcidB{} and Boris Jmailov $^{2}$ }

\AuthorNames{Roman Tkachenko, Vladimir Korchagin and Boris Jmailov}
\AuthorCitation{Tkachenko, 
 R.; Korchagin, V.; Jmailov, B.}
\address{%
$^{1}$ \quad Institute of Physics, Southern Federal University, Stachki Avenue 124, Rostov-on-Don 344090, Russia; vkorchagin@sfedu.ru\\
$^{2}$ \quad Institute of High Technologies and Piezotechnics, Southern Federal University, 10 Milchakova Street, Rostov-on-Don 344090, Russia; bbzhmaylov@sfedu.ru}

\corres{Correspondence: rtkachenko@sfedu.ru}

\firstnote{These authors contributed equally to this work.} 
\abstract{We numerically study the origin of the multi-armed spiral structure observed in the circumnuclear gaseous mini-disks of nearby galaxies. We show that the presence of dust in such disks and its interaction with the gravitationally stable gaseous component leads to the development of a multi-armed spiral structure. As a particular example, we study the formation of the multi-armed spiral pattern in the mini-disk of the galaxy NGC 4736, for which the observational data for the rotation and the density distribution are available. We find that the multi-armed spiral structure grows in the stable gaseous mini-disk of NGC 4736 if the gas-to-dust ratio is about 5–20 percent.  We also demonstrate that together with the dust concentration, the important factor for the development of instability is the size of the dust grains. A nonlinear multi-armed spiral pattern develops in the stable gaseous disk with sizes of grains larger than one micron. If future observations confirm the presence of a large amount of dust in the mini-disks of galaxies, this will pinpoint the mechanism of the formation of the multi-armed spiral structure in them.}

\keyword{physical data and processes : 
hydrodynamics; physical data and processes : instabilities; galaxies : spiral; galaxies : evolution; methods: numerical} 

\begin{document}

%%%%%%%%%%%%%%%%%%%%%%%%%%%%%%%%%%%%%%%%%%

\section{Introduction}

\label{sec:intro}

The first HST observations of central regions of nearby galaxies revealed the existence of dusty gaseous mini-disks with sizes from a few tens to a few hundred parsecs (e.g.,~\cite{Car1,Car2,Car3}).  These observations also demonstrated the presence of rich variety of features in mini-disks, such as rings and multi-armed spirals. \citet{Martini} published a catalog of optical and near-infrared images of the central regions of 123 nearby galaxies. Observations show that the majority of active and inactive galaxies have in their central regions circumnuclear mini-disks with a wide range of structures, spirals, and rings differing by the number of arms, symmetry, and~the pitch angles. Observations of the central regions of elliptical galaxies \citep{Gan} show that cold dusty gaseous disks exist in the central regions of several massive elliptical galaxies as well. Understanding the dynamics of the central mini-disks is important to clarify the mechanism that fuels the central black holes in galaxies and to understand the evolution of the galactic mini-disks themselves.  

To explain the origin of the multi-armed spiral structure in the mini-disks, a few possible mechanisms have been suggested. One of the first possible explanations of the phenomenon was suggested by \citet{Eng}. They supposed that the spiral arms in the circumnuclear disks are related to the grand-design spiral arms, arguing that gas density waves behave differently compared to stellar density waves and are not completely absorbed by the inner Lindblad resonance in galactic disks. 
\citet{Wad} studied the dynamics of the multi-component mini-disks influenced by the potential of a weak mini-bar. \citet{Wad} took into account the self-gravity of gas, as~well as its cooling and heating, and described in their simulations the formation of cusps and filaments on a parsec scale resembling the observed morphological patterns in the circumnuclear mini-disks. \citet{Wong} used high-resolution hydrodynamic simulations to study the properties of nuclear spirals driven by a weak bar-like potential. They found that the amplitude of the spirals increases towards the center of the disk due to a geometric effect developing nonlinear shocks even in a weak mini-bar potential. \citet{Wong} find that induced mass inflow rates are enough to feed the accretion on the central black hole in various types of Seyfert galaxies. \citet{Kim} also investigated the influence of the galactic bar on the spiral structure development by placing the central circumnuclear disk in an ellipsoidal potential. However, in~this bar potential case, the~authors managed to obtain only a two-armed spiral structure in the~mini-disk. 

\citet{Tran} numerically studied the formation of circumnuclear gas structures caused by the tidal disruption of molecular clouds in galactic nuclei. They considered the galactic nuclei to be composed of a supermassive black hole and a nuclear star cluster and simulated the infall and disruption of a molecular cloud within the central parsecs of the galactic nuclei. Their findings indicate that the formation of compact rings of gas occurs in the nuclear regions dominated by the gravity of the nuclear star~cluster.

\citet{Elm1998} discussed possible mechanisms of the formation of nuclear spirals. These authors rejected the origin of the nuclear spirals due to the bar potential, arguing that inner spirals are not as symmetric as dust features in the bars. They also exclude the outer spiral structure as a possible mechanism for the formation of the inner spiral structure arguing that these features are well inside the inner Lindblad resonance and inside the Q-barrier caused by the bulge potential. \citet{Elm1998} also excluded the gravitational instability of the gaseous mini-disks as a possible explanation of mini-spirals, estimating that the value of  Toomre’s Q-parameter in such disks exceeds well the value of Q = 2.  \citet{Elm1998} suggest that the observed spirals might be sound waves that propagate in the gaseous disk of the galaxy towards its center and amplify at small galactocentric radii. This mechanism, however, needs a smooth distribution of gas in a galactic disk close to its center, which is not the case in the central regions of many galaxies.
The discussed mechanisms thus have some weak points and do not account for the available observational~data. 

One more possibility was discussed by \citet{Orlova}. 
It is well known that dust can significantly destabilize the proto-planetary disks so that the admixture of about 2$\%$ of dust can enhance the growth rates of the unstable spiral modes in the proto-planetary disks \citep{noh1992examination}. \citet{Orlova} assumed that dusty circumnuclear mini-disk can behave the same way so that the observed mini-spirals result from the gravitational instability of gaseous mini-disks destabilized by dust.
\citet{Orlova} and \citet{Theis} further developed this idea by modeling the dynamics of dusty gaseous mini-disks. These authors, however, studied the dynamics of the dusty-gaseous minidisks assuming an unrealistically high mass of the gaseous component so that the minimum value of the Toomre Q-parameter \citep{Toomre} of a gaseous component was small enough and the purely gaseous disk was unstable. Moreover, Theis and Orlova \citep{Theis} were able to follow the dynamics of the dusty-gaseous disks during a short time up to 4.5 $\times$ 10$^8$ years, and~for the large concentration of dust no longer than 3 $\times$ 10$^7$ years. The~question remains, therefore, if and under which conditions the dust can destabilize the circumnuclear galactic disks. We address this question at a new level using observationally based density distribution and the rotation curve of the circumnuclear mini-disks. The availability of high-resolution observations of the central regions of nearby galaxies, as well as data on gas distribution, the dust-to-gas ratio in the circumnuclear disks, the~rates of star formation in them, the~rotation curves of mini-disks, and the masses of the central black holes, allow us to discuss the results quantitatively (e.g.,~\cite{Terrazas,Shioya,Gerin}). We choose in this paper the galaxy NGC 4736 as an observational basis to study the dynamics of the circumnuclear disks. Similar studies of the dust influence on the stability of the gas--dust disks were carried out by \citet{Tom1,Tom2}, but~in the context of circumstellar~disks.

As was stressed by \citet{ChanHerov}, the~problem of stability is of great relevance to studying Newtonian and general-relativistic objects. \citet{Her} discussed the dynamics of a non-adiabatically collapsed spherically symmetric fluid taking into account dissipative effects destabilizing the fluid. In~this paper, we discuss in here the dynamics of a non-relativistic gravitating multi-component disk.

Section~\ref{sec:obs} describes the available observational data for the circumnuclear disk in NGC 4736.
Section~\ref{sec:model} discusses the model we adopted to model the dynamics of a gas--dust mini-disk. Section~\ref{sec:Res} describes the parameters of the mini-disk of the galaxy NGC 4736, as~well as the results of modeling the dynamics of the gas--dust mini-disk of this galaxy. Section~\ref{Conclusion} summarizes the results of our~study.

\section{Observational~Data}\label{sec:obs}

NGC 4736 (Messier 94) is an early type spiral galaxy of the type (R)SA(r)ab \citep{Korm} with a low ionization nuclear emission region. The~central part of the galaxy has a massive circumnuclear dusty gaseous disk with a multi-armed spiral structure not related to the large-scale spiral structure of the galaxy. The~nuclear spirals are not associated with star formation and they are very irregular with both trailing and leading components that often cross. The~number of the arms, traced by the dust, grows from five in the central regions of the mini-disk to approximately twenty at the disk's periphery \citep{Elmegreen1}. 

The HST images of the circumnuclear mini-disk of NGC 4736 with a developed multi-armed spiral structure can be found in \citet{Kim} and \citet{Elmegreen1}.  The~spiral pattern in the mini-disk of NGC 4736 is nonlinear with arm/interarm density variations in the gaseous disk of about one hundred percent \citep{Elmegreen1}.

\subsection{Gas and Dust in the Circumnuclear Disk of NGC~4736}
\label{sec:gasdust}
 Gas in the mini-disk of NGC 4736 is almost entirely in a molecular phase (90--100$\%$) inside the central kiloparsec of the galaxy \citep{Gerin}.
Shioya~et~al.~\cite{Shioya}, Van der Laan~et~al.~\cite{Laan}, and Muñoz-Tuñón~et~al.~\cite{Mu1} estimated the gas surface density in central regions of the circumnuclear disk of NGC 4736 to be about 120 M$_{\odot}$pc$^{-2}$, decreasing exponentially with a radial scale length of about 0.5 kpc.  The~velocity dispersion of the gas is about 6--8~km$\cdot$s$^{-1}$~\citep{Anahi,Wong,Shioya}. Similar to \citet{Wong}, we use in our simulations the value of  7~km$\cdot$s$^{-1}$. The~galaxy NGC 4736 is observed nearly face-on, so the data on the thickness of the dusty mini-disk of the galaxy NGC 4736 is absent. We accept in this paper the thickness of the circumnuclear disk of NGC 4736 to be 100 pc according to \citet{Gerin}. This choice is reasonable because the typical vertical scale height of molecular disks varies between 20 and 200 pc \citep{Patra}.

Observational estimates of the mass fraction of dust in the mini-disk of the galaxy NGC 4736 are absent. \citet{Mu2009,Mu2011} estimated the oxygen abundance in the central regions of the galaxy to be $(O/H)=9.96$ dex. 
The Sun's oxygen abundance is $(O/H)_{\odot}=8.69$ dex \citep{asplund}. Then the relative abundance of oxygen $[O/H]={\log_{10}(O/H)}-\log_{10}(O/H)_{\odot}$ in the central parts of NGC 4736 is equal to 1.27 dex (in the case IMF is assumed from \citet{K01}), which means that the metallicity in the central part of NGC 4736 about 18~times higher than in the Sun. Assuming that the dust density has a linear dependence on metallicity, the~dust-to-gas ratio in the circumnuclear disk of NGC 4736 can reach \mbox{10--20$\%$~\citep{Mu2009}}. This estimate is in line with studies of the dust-to-gas ratio in the central parts of early type galaxies \citep{kokusho2019dust}. 
The size distribution of the dust particles is an open question in the circumnuclear disks. The~typical size of dust particles in the galactic interstellar medium is about 0.1--1 $\upmu$m \citep{Draine}. However \citet{Maiolino1,Maiolino2} found that the properties of dust particles in the circumnuclear regions of galaxies may differ from those in the diffuse ISM of our galaxy, most of the dust in the central regions of galaxies is contained in grains with sizes larger than 3 $\upmu$m. We take this into account in our~simulations.

\label{sec:data}

\subsection{Mini-Disk Rotation Curve and the Central Black~Hole}
\label{sec:RC}

\begin{figure}[H]
\includegraphics[width=0.40\hsize]{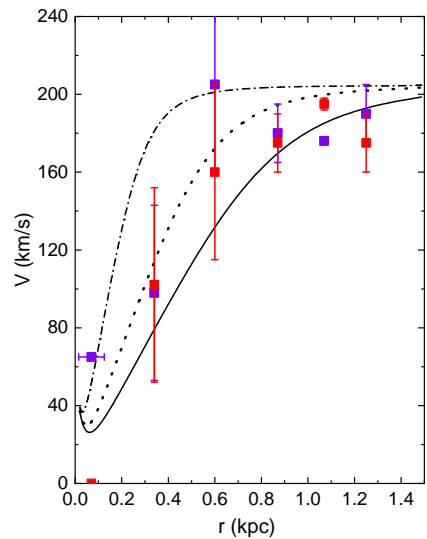}
\caption{The rotation curve of the mini-disk of galaxy NGC 4736. The~observational data are based on the rotational lines of CO (J = 1-0 transition: red, and~\mbox{J = 2-1} transition: purple squares) taken from~\citep{Gerin}. Solid, dotted, and dash-dotted lines are the rotation curves adopted in this paper.
\label{Rot}}
\end{figure}

Figure~\ref{Rot} shows the observational data for the rotational velocity of molecular gas in the circumnuclear disk of NGC 4736, obtained from the rotational lines of CO (J = 1-0 and \mbox{J = 2-1}) \citep{Gerin}.  As~one can see from the figure, measurements have significant uncertainties for the rotational velocity as well as in the radial scale length of transition from solid body rotation to the flat rotation curve with the velocity of 200 km$\cdot$s$^{-1}$ \citep{Gerin,Wong,Bratek}. Therefore, in~our models, we use three rotation curves that cover approximately the uncertainties in measurements.
The central region of NGC 4736 has a supermassive black hole with mass $M_{BH}=6.8\cdot10^6 M_{\odot}$ \citep{Terrazas} which was taken into account in constructing the theoretical rotation curve in our~simulations.

We use available observational data to build the theoretical rotation curves of the mini-disk of the galaxy NGC 4736 shown in Figure~\ref{Rot}. The~stellar bulge and the dark matter halo determine the main contribution to the equilibrium rotation of the mini-disk and are taken into account in the equilibrium model. The~theoretical rotation curve is described in more detail in Section~\ref{sec:model}.

\section{The~Model} 
\label{sec:model}
We build the initial axisymmetric distribution of gas and dust in the mini-disk of the galaxy taking into account the observed parameters. The~distribution of the surface density of gas and dust in the disk is given by the equation:
\begin{equation}\label{eq:DensDistr}
 \sigma_{g,d}(r)=\sigma^0_{g,d}\exp(-r/r_s)  \,,
\end{equation}
where $\sigma_{g,d}$ is the surface density, the~indices ``\emph{g}'' and ``\emph{d}'' stand for the gas and the dust components, the~radial scale length of the disk, $r_{s}$, equal to 0.5 kpc, is taken from observations~\citep{Shioya}. The~density of gas in the central regions of NGC 4736, $\sigma^0_{g}$, equal to 120 $M_{\odot}~\text{pc}^{-2}$, which is taken from \citet{Laan}. Then the total mass of gas in the mini-disk of NGC 4736 is equal to 1.58$\cdot10^8 M_{\odot}$. The~dust distribution is determined by the dust-to-gas ratio $\epsilon=\sigma_{d}/\sigma_{g}$.
In this paper, we consider three values for this ratio $\epsilon$ = 0.05, 0.1, and 0.2. The~thickness of the molecular disk $h$ is taken according to \citet{Gerin} to be 0.1~kpc.

The dynamics of the mini-disk are described by a set of multicomponent hydrodynamical equations, which we solve with help of the two-dimensional multicomponent numerical code based on ZEUS 2D, which is described in detail in \citet{Stone}. The~equations are solved on a logarithmic Eulerian grid in polar coordinates with a grid resolution of \mbox{256 $\times$ 256}, which provides high enough accuracy in the inner parts of the disk. The~inner boundary of the computational domain is taken at $r_{in}= 0.01$ kpc, and~the outer $r_{out}= 1.5$ kpc. The~accretion onto the central black hole is taken into account at the inner boundary, so that the mass of gas and dust, flowing through the inner boundary, is added to the mass of the central black hole in the our~case. 

In our simulations, we use dimensionless units with the gravitational constant equal to one, the unit of length equal to 1 kpc, and~the unit of mass equal to $10^9M_{\odot}$. With~these units, the~unit of time is equal to $1.49 \cdot 10^{7}$ years, the~velocity unit is  65.6 km$\cdot$s$^{-1}$, and~the surface density unit is $10^3$\emph{M}$_{\odot}$~pc$^{-2}$.

The behavior of the two-component dusty gaseous mini-disk is described by the set of continuity equations, equations of motion, and Poisson's equations in polar coordinates for the gas ``\emph{g}'' and dust ``\emph{d}'' components:
\begin{equation}\label{eq:Continuity}
    \frac{\partial{ \rho}_{g,d}}{\partial t} +
         \nabla \cdot (\rho_{g,d} \cdot \vec{v}_{g,d}) = 0 \,\,\, ,
\end{equation}
here, $\rho_{g,d}$ are the volume densities of gas and dust, and $\vec{v}_{g,d}$ are the velocity vectors of the gas and dust components.
The equations of motion have the forms:
\begin{equation}\label{eq:Dyn1}
   \frac{\partial{\vec{v}_g}}{\partial t} + (\vec{v}_g \cdot \nabla) \vec{v}_g
         =- \frac{\nabla P_g}{\Sigma_g} 
         - \nabla (\Phi + \Phi_{\rm H})
         - \epsilon\frac{(\vec{v}_g-\vec{v}_d)}{t_{stop}} \,\,\, , 
\end{equation}
\begin{equation}\label{eq:Dyn2}
   \frac{\partial{\vec{v}_d}}{\partial t} + (\vec{v}_d \cdot \nabla) \vec{v}_d
         = - \nabla (\Phi + \Phi_{\rm H})
         + \frac{(\vec{v}_g-\vec{v}_d)}{t_{stop}} \,\,\, . 
\end{equation}

Equations (\ref{eq:Dyn1}) and (\ref{eq:Dyn2}) are Eulerian-type equations describing the behavior of dust and gas components taking into account the gaseous pressure $P_g$, the~self-gravity of the multi-component disk $\Phi$,  the~gravitational force from the external stationary potential $\Phi_H$ of the galactic bulge, halo, and the central black hole. The~dust dynamics are described by the pressureless Equation~(\ref{eq:Dyn2}). The~last terms in Equations~(\ref{eq:Dyn1}) and (\ref{eq:Dyn2}) describe the friction force between the two components. 
Following \citet{Cha,Zhu,Vorob}, we choose the friction force to be proportional to the velocity difference between gas and dust, $\epsilon=\frac{\rho_d}{\rho_g}$ is the dust-to-gas ratio. The~value of $t_{stop}$ is a characteristic time of momentum exchange between the dust and the gas components which according to \citet{Stoyanovskaya,Stoya} can be expressed by the equation:
\begin{equation}\label{eq:tstop}
   t_{stop}=\frac{a\rho_{in}}{\rho_{g}c_s}  \,\,\, , 
\end{equation} 
where $a$ and $\rho_{in}$ are the radius and the density of a dust particle accepted to be 3.8 g$\cdot$cm$^{-3}$ for silicate grains \citep{Draine}. We consider three different sizes of dust particles equal to 0.1, 1, and 5 $\upmu$m. The~last parameter determining the behavior of a gaseous mini-disk is the sound speed in the gaseous component $c_s$ which is assumed to be equal to 7 km$\cdot$s$^{-1}$.
It should be noticed that the expression for $t_{stop}$ can be represented by Equations~(\ref{eq:tstop}) if the dust sizes ``\emph{a}'' are less than the mean free path of the gas molecules ``$\lambda$'' in the circumnuclear disk when $a<2.25\lambda$ (Epstein regime \citep{Eps,Good} for gas flowing around a rigid body) \citep{Stoyanovskaya}.

Equations~(\ref{eq:Continuity})--(\ref{eq:Dyn2}) are solved using the operator-split solution procedure as described in \citet{Stone}, which is split into the transport and source steps. In~the transport step, advection is performed by a second-order Van Leer interpolation scheme \citep{Leer}. To~take into account the friction forces in the source step we use the explicit approximation scheme from \citet{Stoyanovskaya,Stoya}, the~stability of which (along with the Courant--Friedrichs--Levy criterion) is characterized by the condition $dt<\frac{2t_{stop}}{1+\epsilon}$. We also performed calculations with a semi-implicit (mixed-layer) scheme \citep{Stoyanovskaya,Stoya}, which showed identical results. The~use of explicit and semi-implicit schemes is described and investigated in detail in \citet{Vorob,Cha,Stoyanovskaya,Stoya}.

The values $\Phi$ and $\Phi_{\rm H}$ denote the potential of the mini-disk disk and the external stationary potential of the galactic bulge, halo, and~the central black hole, respectively. Potentials of gas and dust disks are solved by applying the two-dimensional Fourier convolution theorem in polar coordinates which is described in \citet{press1996}. The~Poisson equation can be represented as:
\begin{equation}\label{eq:Poisson}
  \Delta \Phi = 4 \pi G \left( \rho_g + \rho_d \right) \,\, .
\end{equation}

The system of hydrodynamic equations is closed by the equation of state:
\begin{equation}\label{eq:Pressure}
    P_g = K \rho_g^{\gamma} \,\,\, ,
\end{equation}
where $\gamma$ is the polytropic index of the gas, equal to one for an isothermal gas. The~constant $K$ is determined by taking the sound speed of the gas in the mini-disk equal to $c_s$~=~7~km$\cdot$s$^{-1}$~\citep{Wong} from the expression:
\begin{equation}\label{eqstate}
    c_s = \sqrt{\frac{\partial{P_g}}{\partial \rho_g}} =\sqrt{\gamma K \rho_g^{\gamma-1}} \,\,\, .
\end{equation}

We model the observed rotation curve of the galaxy NGC 4736 using the equation:

\begin{equation}\label{eq:rc1}
   v_{\rm rot}(r) = \sqrt{\frac{GM_{BH}}{r}    
   +v_{\rm c}^2+ r\frac{\partial{\Phi}}{\partial r}}  \,\,\, .
\end{equation}

The first term in Equation~(\ref{eq:rc1}) is the contribution to the rotation curve by the central black hole. We accept according to \citep{Terrazas} the mass of the central black hole to be equal to \mbox{$M_{BH}=6.8\cdot10^6 M_{\odot}$}.
The second term in Equation~(\ref{eq:rc1}) is the contribution to the rotation curve of the external stationary axisymmetric potential of the galactic bulge and the dark matter halo and can be described using the equation taken from \mbox{\citet{Shioya}}:
\begin{equation}\label{eqvcirc}
   v_{\rm c}(r) = v_{\rm 0} \cdot \frac{{r}}
    {\left[ (r_{\rm 0})^{4} + \left( 
                 {r}\right)^{4} \right]^{1/4}}  \,\,\, .
\end{equation}

\citet{Shioya} used Equation~(\ref{eqvcirc}) to model the rotation curve of the mini-disk of the galaxy NGC 4736 assuming that the mini-disk rotation is determined by the external spherically symmetric stationary potential of the galactic bulge and halo. To~build the equilibrium rotation curve of the mini-disk, we also take into account the mini-disk’s self-gravity in the last term of Equation~(\ref{eq:rc1}). We note here that the influence of the mini-disk self-gravity is less than two percent compared to the halo potential, and~the central black hole affects the rotation curve of the mini-disk closer than 50 pc to the galactic center.

To fit observational data, the~value of $v_{\rm 0}$ in Equation~(\ref{eq:rc1}) was accepted to be equal to 200~km$\cdot$s$^{-1}$. The~value $r_{\rm 0}$ is the characteristic scale length that determines the transition from solid body rotation. In~this paper, we assume three different values for this parameter equal to 0.3, 0.6, and 0.9 kpc. Figure~\ref{Rot} shows the theoretical rotation curves for different values of $r_{\rm 0}$: 0.3 kpc (dash-dotted line), 0.6 kpc (dashed line), and~0.9 kpc (solid line). The~theoretical rotation curves shown in Figure~\ref{Rot} were calculated for a gas-to-dust mass ratio $\epsilon$ equal to ten percent. The~rotation curves do not change much for other accepted values of dust-to-gas ratios.
The radial velocities of gas and dust components at the initial moment of time in the simulation are assumed to be~zero.

\begin{figure}[H]
\includegraphics[width=0.45\hsize]{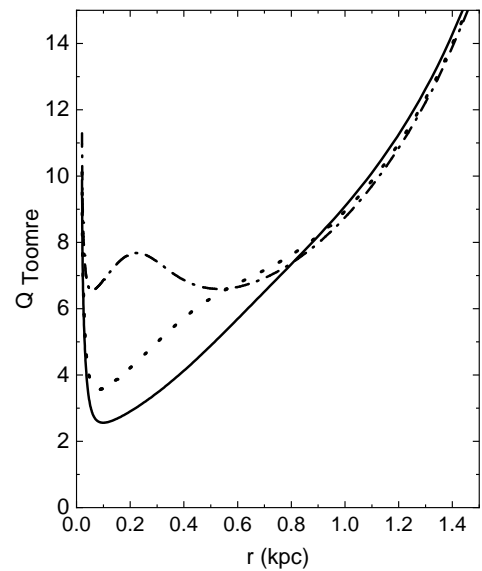}
\caption{Radial dependence of the Toomre Q-parameter in the circumnuclear disk of NGC 4736 for three different initial rotation curves with $r_{\rm 0}$ as 0.3 kpc (dash-dotted line), 0.6 kpc (dash line), and~0.9~kpc (solid line).
\label{fig-VR} }
\end{figure}

Toomre's Q-parameter \citep{Toomre} for the gaseous component of the mini-disk determining the growth of small-scale ring perturbations can be written as:
\begin{equation}\label{Qpar}
   Q=  \frac{c_s\kappa}{\pi\sigma_gG}\,\,\, ,
\end{equation}
where $\kappa$ is the epicyclic frequency of the gaseous mini-disk.  Figure~\ref{fig-VR} presents the radial dependence of the Q-parameter for three different equilibrium rotation curves with $r_{\rm 0}$, 0.3 kpc (dash-dotted line), 0.6 kpc (dashed line), and~0.9 kpc (solid line). As~one can see from Figure~\ref{fig-VR}, the~minimum values of the Q-parameter are larger than 2.58 in all cases, which demonstrates that the purely gaseous mini-disk of NGC 4736 is stable. This was also supported by \citet{Kenn}, who examined fifteen galaxies and found empirically that the critical value of the Q-parameter was $Q_{c}$ = 1.59. The~stability of the purely gaseous mini-disk
of NGC 4736 was confirmed also in our numerical~simulations.

\vspace{-3pt}

The growth of perturbations in the gaseous or dust components of the mini-disk is measured with the help of the global logarithmic Fourier amplitudes determined as:
\begin{equation}\label{Amplitude}
     A_m = \log_{10}( {1 \over {M_{\rm s}}}
     \left| \displaystyle \int_{0}^{2 \pi} \int_{R_{\rm in}}^{R_{\rm ex}}
     \Sigma(r,\phi) r dr \, e^{-im\phi} d\phi \right|) \,\,\,\,\,,
\end{equation}
where $M_{s}$ is the mass of the gas or dust components enclosed in the region between $R_{in}=10$ pc and $R_{in}=250$ pc. The~value of 250 pc was chosen empirically based on the characteristic radius of the existence of a spiral structure in the circumnuclear disk of the galaxy NGC 4736. $A_m$ is the logarithmic Fourier amplitude for a given mode m.
The initial amplitudes of the perturbations in the gaseous and dust components were set by the random addition of density perturbations of relative amplitude $10^{-5}$.

The parameters of simulated models are listed in Table~\ref{table:tab}. The~first column shows the names of the models, the~second shows the values of the dust-to-gas ratio $\epsilon$. The~size of the particles $a$, determining the characteristic time of the momentum exchange between the two components is given in the third column, and~scales of the rotation curves $r_0$, which also characterize the distribution of Q-parameter, are given in fourth~column.

\begin{table}[H]
\caption{The parameters of the circumnuclear disk of NGC 4736 used in our~simulations.\label{table:tab}}

		\begin{tabular}{cccc|cccc}			\toprule
			\textbf{Model} & \boldmath{$\epsilon$} & \boldmath{$a$}&\boldmath{ $r_0$}  & \textbf{Model} & \boldmath{$\epsilon$} & \boldmath{$a$ }& \boldmath{$r_0$}\\
            \midrule
         M111  &  0.05 & 0.1 $\upmu$m & 0.3 kpc & M223  & 0.1 & 1.0 $\upmu$m & 0.9~kpc \\ 
         M112  &  0.05 & 0.1 $\upmu$m & 0.6 kpc & M231  & 0.1 & 5.0 $\upmu$m & 0.3~kpc  \\
         M113  &  0.05 & 0.1 $\upmu$m & 0.9 kpc & M232  & 0.1 & 5.0 $\upmu$m & 0.6~kpc  \\ 
         M121  &  0.05 & 1.0 $\upmu$m & 0.3 kpc &  \textbf{M233}  & \textbf{0.1} & \textbf{5.0 $\upmu$m} & \textbf{0.9 kpc}  \\
         M122  &  0.05 & 1.0 $\upmu$m & 0.6 kpc & M311  & 0.2 & 0.1 $\upmu$m & 0.3~kpc  \\
         M123  &  0.05 & 1.0 $\upmu$m & 0.9 kpc & M312  & 0.2 & 0.1 $\upmu$m & 0.6~kpc  \\
         M131  &  0.05 & 5.0 $\upmu$m & 0.3 kpc & M313  & 0.2 & 0.1 $\upmu$m & 0.9~kpc  \\
         M132  &  0.05 & 5.0 $\upmu$m & 0.6 kpc & M321  & 0.2 & 1.0 $\upmu$m & 0.3~kpc  \\
         M133  &  0.05 & 5.0 $\upmu$m & 0.9 kpc & M322  & 0.2 & 1.0 $\upmu$m & 0.6~kpc  \\
         M211  & 0.1 & 0.1 $\upmu$m & 0.3 kpc   & M323  & 0.2 & 1.0 $\upmu$m & 0.9~kpc  \\
         M212  & 0.1 & 0.1 $\upmu$m & 0.6 kpc   & M331  & 0.2 & 5.0 $\upmu$m & 0.3~kpc  \\
         M213  & 0.1 & 0.1 $\upmu$m & 0.9 kpc   & M332  & 0.2 & 5.0 $\upmu$m & 0.6~kpc  \\
         M221  & 0.1 & 1.0 $\upmu$m & 0.3 kpc   & M333  & 0.2 & 5.0 $\upmu$m & 0.9~kpc  \\
         M222  & 0.1 & 1.0 $\upmu$m & 0.6 kpc   &    &   &    &    \\
       
			\bottomrule
		\end{tabular}
	
	\noindent{\footnotesize{Here, $\epsilon$ is the gas-to-dust ratio, $r_0$ is the radial scalelength of the rotation curve, and {a} is the size of a dust particle. The~reference model is highlighted in bold and discussed in detail in Section~\ref{sec:Res}.}}
	\label{obsdata}
\end{table}

It should be noticed that in this paper we do not consider possible star formation processes in the circumnuclear disk. We also do not take into account the influence of the stellar component, assuming that its influence is taken into account in the external potential. We do not consider also the processes of the growth and the destruction of the dust particles and distribution of dust grains by~size. 
\section{Results and~Discussion}
\label{sec:Res}

The model \textbf{M233} in Table~\ref{table:tab} was selected as a reference for two reasons: (i) the model parameters are reasonable and in agreement with the observational values; (ii)~based on the comparison of the theoretical spiral pattern developed in the dusty-gaseous mini-disk with the spiral structure observed in the circumnuclear disk of NGC~4736.

\begin{figure}[H]
\widefigure{
	\includegraphics[width=0.70\textwidth]{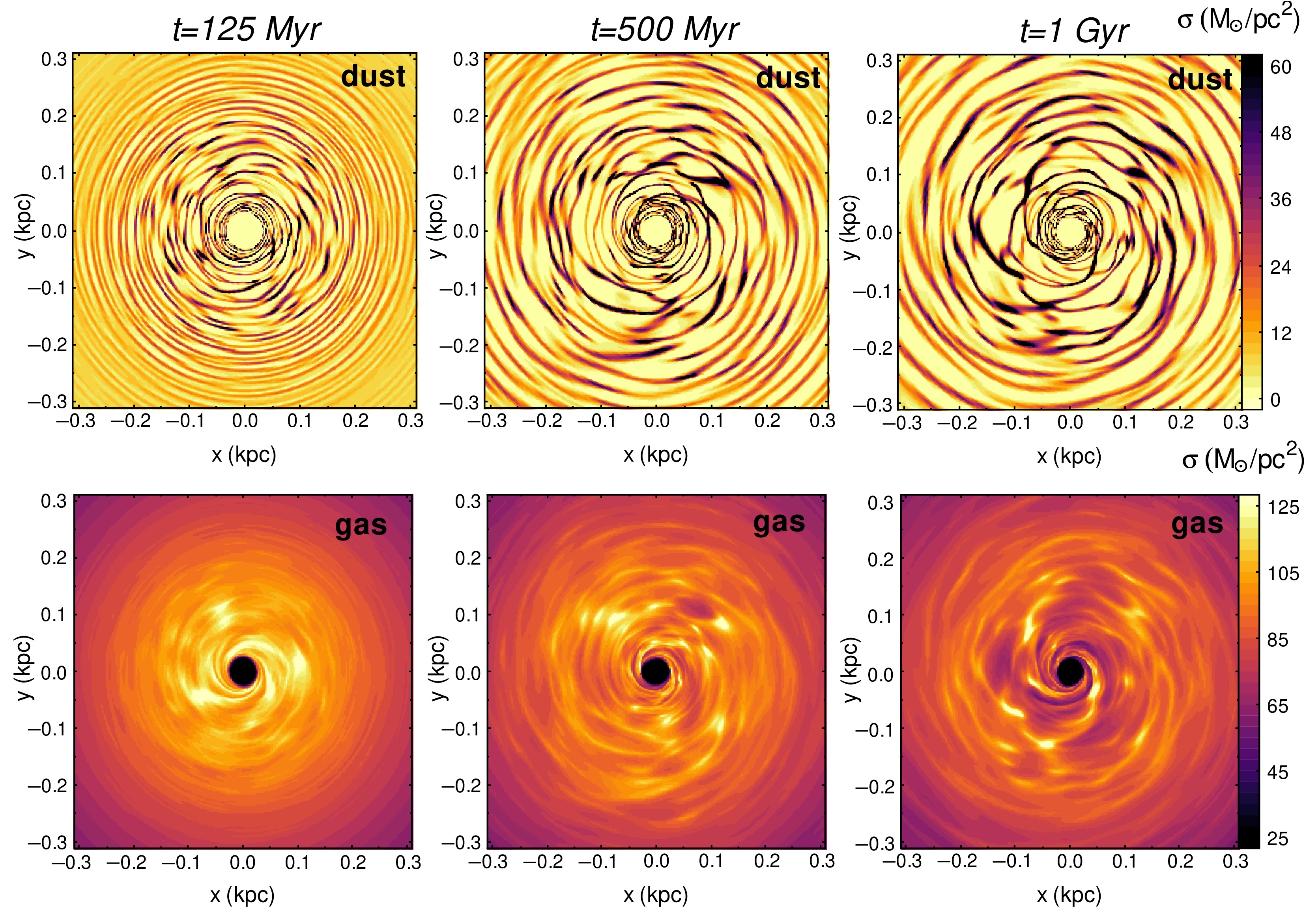} }
	\caption{Snapshots of the temporal evolution of dust (upper row) and gas (lower row) surface densities in the reference model \textbf{M233} 
taken at times $t=125$ Myr, $t=500$ Myr, and~$t=1$ Gyr (from left to right). Surface density is measured in [\emph{M}$_{\odot}$\text{pc}$^{-2}$]). The~instability of the gaseous dusty disk results in the development of the nonlinear multi-armed spiral structure.
\label{fig-Dens1} }
\end{figure}

In this model, the~minimum value of the Q-parameter in the gaseous disk is equal to 2.58, so the purely gaseous disk is stable which was confirmed by the numerical simulations. A~ten percent admixture of dust strongly destabilizes the disk.
Figure~\ref{fig-Dens1} presents snapshots of the temporal evolution of perturbations taken at $t=125$ Myr, $t=500$ Myr, and $t=1$ Gyr.
As can be seen from the figure, during~the evolution of perturbations, a~multi-armed spiral structure develops in the dusty gaseous disk. The~spiral structure stabilizes quite quickly, and~at time $\sim$100 Myr is already noticeable. The~spiral arms in the dust component correlate in general with those of the gaseous disk. The~spiral arms, developing in the dust, differ, however, by~a large density contrast between the arm and the inter-arm regions and the presence of a larger number of small-scale armlets. At~larger radii, the~spiral arms in the dust component become more tightly wound; however, such behavior is not traced for the gaseous spirals. It should be noticed that the nonlinear spiral structure exists in the dusty gaseous disk for quite a long time, at~least during one~Gyr.

\nointerlineskip

\begin{figure}[H]
\includegraphics[width=0.63\textwidth]{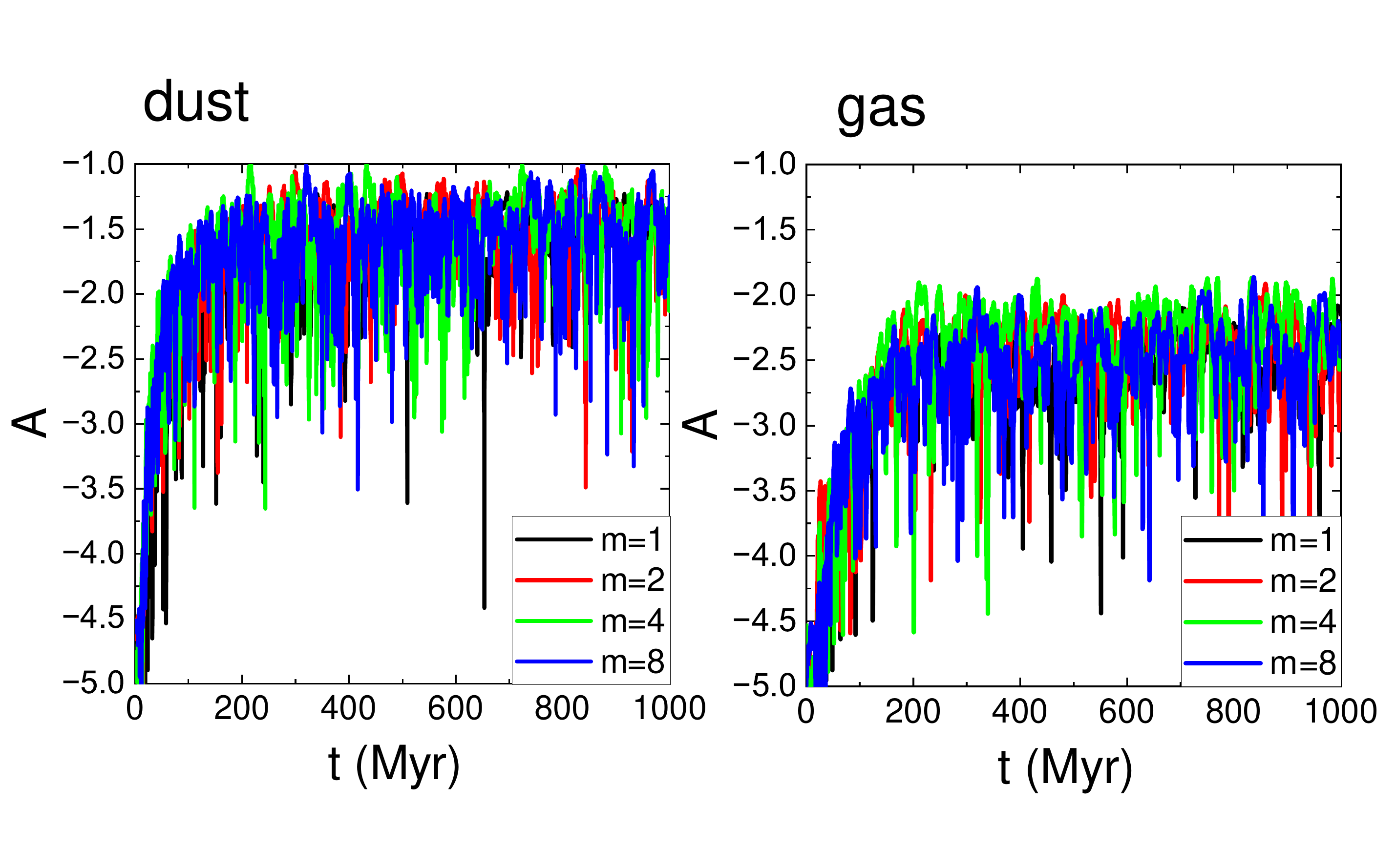}
\caption{Evolution of global Fourier amplitudes of dust (on the \textbf{left}) and gas (on the \textbf{righ}t) components in the reference model \textbf{M233}, for~the modes: m = 1---black line, m = 2---red, m = 4---green, and m = 8---blue.
\label{fig-Glob}}
\end{figure}

Figure~\ref{fig-Glob} shows the evolution of the logarithmic global Fourier amplitudes (see \mbox{Equation~(\ref{Amplitude}})) in the dust (left) and in the gaseous (right) components for the reference model for the Fourier modes m = 1, 2, 4,  and~8. As~one can see from the figure, the~growing perturbations in the gaseous and the dust subsystems reach the nonlinear saturation stage after about 100--150 million years. We find that perturbations with any number of spiral arms (10 harmonics were studied) grow at approximately the same rate and saturate at the level of a few percent in the dust, and~less than one percent in the gaseous component. However, due to an interplay of a large number of harmonics, the~response in the gas and in the dust components is highly~nonlinear.

\begin{figure}[H]
\includegraphics[width=0.63\textwidth]{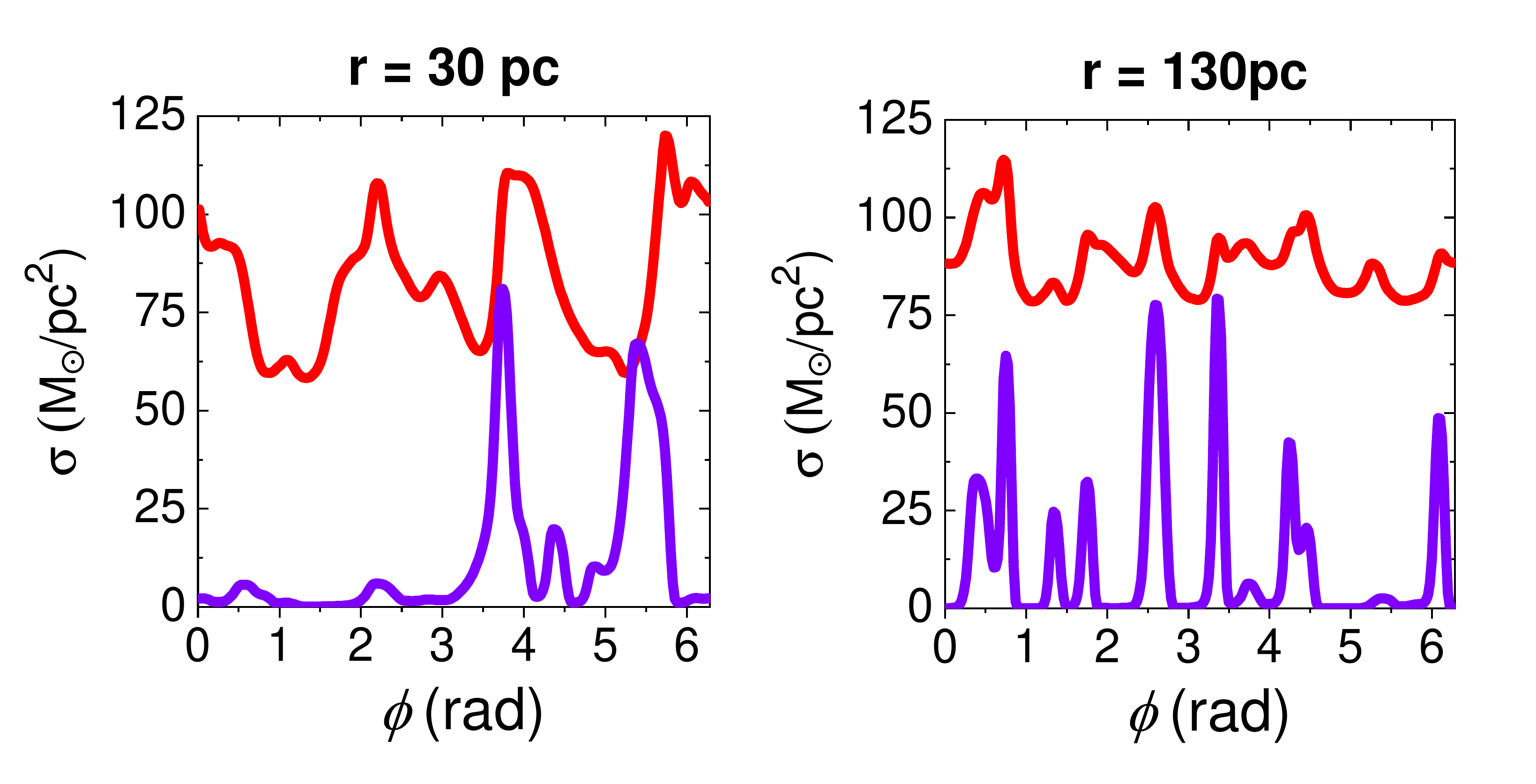}
\caption{Azimuthal dust (purple color) and gas (red color) density variations (in [\emph{M}$_{\odot}$~pc$^{-2}$]) for the reference model \textbf{M233} at the two fixed radii: \emph{r} = 30 pc (on the \textbf{left}) and \emph{r} = 130 pc (on the \textbf{right}) at the time $t=1$ Gyr. 
\label{fig-asim} }
\end{figure}

Figure~\ref{fig-asim} shows the azimuthal density variations at a radius of 30 (left) and 130 (right) pc for the reference model \textbf{M233} taken at time 1 Gyr. As~can be seen from the figure, the~arm/interarm density contrast reaches about one hundred percent both in the gaseous and in the dust~components.

\vspace{-6pt}

The peaks of the density perturbations in the dust surface density distribution correlate with the gas density variations. Dust density variations, however, are significantly greater compared to those in the gaseous disk. The~amplitudes of the density variations in the dust become smaller when decreasing the size of the dust particles, i.e.,~with an increasing rate of momentum exchange between the gas and the dust. As~the radius increases from 30~parsecs to 130, the~number of arms increases from 3 to about~8.

\begin{figure}[H]
\widefigure{
	\includegraphics[width=0.60\textwidth]{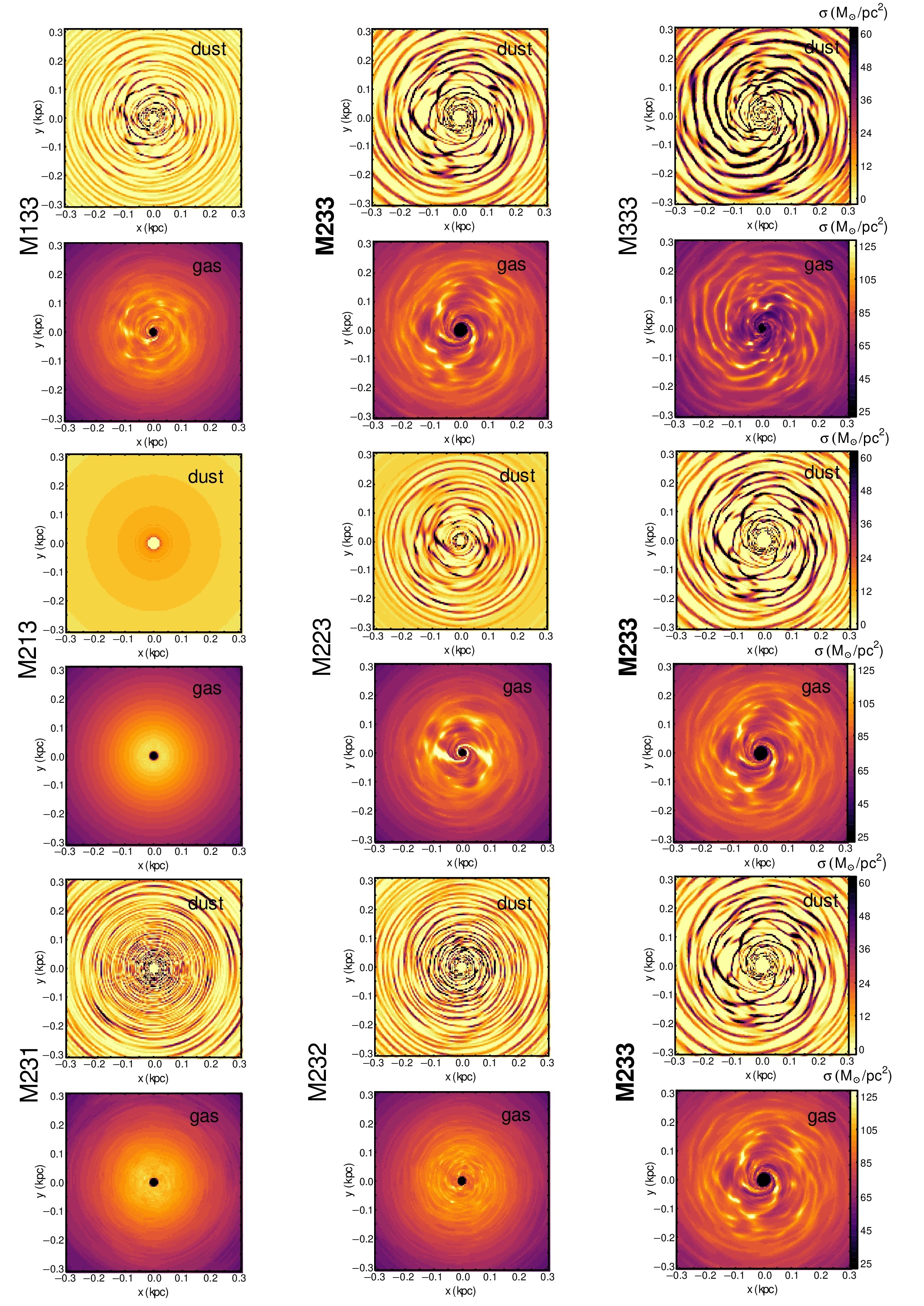} }
	\caption{Distribution of dust (upper rows) and gas (lower rows) surface densities of the different models for which the parameters vary relative to the reference model \textbf{M233} at time $t=1$ Gyr. The~color indicator with surface density (measured in [\emph{M}$_{\odot}~$pc$^{-2}$]) is shown in the figure. 
\label{fig-Dens2} }

\end{figure}

Figure~\ref{fig-Dens2} illustrates the dependence of the spiral pattern developing in the mini-disk on the parameters of dust, and~on the rotation curve of the mini-disk. The~upper two rows show the dependence of the developing spiral pattern on the varying dust content in the fiducial model \textbf{M233} with $r_0$ = 0.9 kpc, $\epsilon$ = 0.1 and a = 5 $\upmu$m (``low'' momentum exchange between dust and gas). As~one can see, an~increase of the dust-to-gas ratio from $\epsilon$ = 0.05 (upper left frames) to 0.2 (upper right frames) leads to the growth of the area involved into the star-gas instability together with the growth of density variations in the arm/interarm regions and the number of armlets developing in the~mini-disk.

\nointerlineskip

The central two rows in Figure~\ref{fig-Dens2} (models M213, M223, and \textbf{M233}) illustrate the dependence of the developing spiral pattern on the size of the dust particles. From~left to right, the size of the dust grains varies from a = 0.1 $\upmu$m (middle left frames ) to 1 $\upmu$m and~5 $\upmu$m (middle right frames). As~Figure~\ref{fig-Dens2} illustrates, a~decrease of the momentum exchange between the dust particles and gas leads to the development of instability in the gas--dust disk, and~to the development of a multi-armed spiral pattern in it. For~the M213 model (with ``strong friction''), the~gas and dust disks remain stable. This can be understood in terms of the stability criterion for a gas--dust medium. In~the case of the limit of infinite drag/perfect coupling between the dust particles and gas (i.e., in the case of the small-sized dust grains), gas and dust behave like one medium, so that Toomre's Q-parameter is determined by the equation \citep{Vorob,Laib}:
\begin{equation}
Q_{T}={c_d \kappa \over \pi G (\Sigma_ g+\Sigma_d)}.
\label{ToomreQ}
\end{equation}

Here, $c_{\rm d}=c_{\rm s}/\sqrt{1+\epsilon}$ is the sound speed in the dust--gas medium, and~$\epsilon$ is dust-to-gas ratio.
Toomre's criterion for strongly coupled gas--dust model M213 gives a minimum value of Toomre's Q-parameter equal to 2.23, pointing to the stability of the mini-disk. This result was confirmed by our numerical simulations (see left frames of the middle row in Figure~\ref{fig-Dens2}). 

Increasing the size of the dust particles leads to a slower momentum exchange between the dust particles and the gas causing, in~turn, the~development of the viscous gravitational instability \citep{Tom1,Tom2} and the growth of the multi-armed spiral structure in the disk. Correspondingly, the~arm/interarm density contrast also grows while increasing the size of the dust~particles.  

The bottom rows in Figure~\ref{fig-Dens2} illustrate the dependence of the morphology of growing spiral perturbations on the parameters of the rotation curve of a mini-disk. The~scale length of the rotation curves varied in our simulations from $r_0$ = 0.3 kpc (left bottom frames) to 0.6 kpc (middle frames) and 0.9 kpc (right bottom frames). As~one can see from the figure, a~shorter scale length of the rotation curve leads to a degeneration of the growing spiral structure and the development of the ring-like perturbations close to the central regions of the mini-disk. We also note that the perturbations in the gaseous component of a mini-disk have small amplitudes and are barely noticeable, contradicting the available observational~data.

Thus, the~development of instability takes place more efficiently when the dust-to-gas ratio and the size of the dust particles reach large values. Not all models are shown in Figure~\ref{fig-Dens2}, but~all the remaining models in Table~\ref{table:tab} follow the behavior described~above.

Strictly speaking, the~dynamics of dusty circumnuclear disks should be considered as a three-dimensional problem so that the dust component, which is dynamically differentiated from the gas, should be more concentrated in the plane compared to the gaseous disk. The~vertical scale height of the molecular gas in the inner galaxy is about 48--160 pc \citep{Nakan}. One can expect that the vertical scale height in the dusty circumnuclear disks is comparable to this value. With~radius of the circumnuclear disk in NGC 4736 about 1 kpc, the~ratio of the width of the circumnuclear disk of NGC 4736 to its radius is about 0.1, which is less than the width-to-radius ratio in the protoplanetary disks, which are often considered in two-dimensional models \citep{Vorob,Stoya,Stoyanovskaya}.
Note also that in the case of the vertical stratification of dust and gas in the circumnuclear disks, dust probably interacts with a smaller amount of gas, and~the destabilizing effect of dust should be stronger. 
However, a~more accurate answer requires a three-dimensional~model.

\section{Conclusions}
\label{Conclusion}

In this work, we studied the influence of the dust component on the evolution of the circumnuclear gaseous dusty disks. Attention was paid to the question of whether the presence of dust in the galactic mini-disks can destabilize their otherwise stable gaseous components and form in them the multi-armed spiral structure observed in the mini-disks of nearby galaxies. As~an example, we modeled the dynamics of the dusty gaseous mini-disk of the galaxy NGC 4736 taking into account the interaction between the components both gravitationally and via the momentum exchange between the subsystems. The~equilibrium models for the mini-disk of this galaxy were based on the available observational data. 
We show that a dust-to-gas ratio in the mini-disk of 5--20$\%$ affects the dynamics of the circumnuclear disk. We find that the admixture of dust in the dusty-gaseous mini-disk leads to its significant destabilization caused by the gravitational instability and to the formation of a highly nonlinear multi-armed spiral structure in the mini-disk. The~multi-armed spiral structure is formed both in the dust and in the gaseous subsystems.
The nonlinear stabilization of instability which replaces the stage of exponential growth of perturbations occurs after about 100 million years and lasts for more than one Gyr (the time that the~simulations were performed). 
The resulting patchy spiral structure is seen both in the gaseous and in the dust subsystems and is highly nonlinear with arm-to-interarm density contrast of about a few hundred percent. In~agreement with observational data, the~spiral armlets developing in the dust subsystem have greater density contrast of arm/interarm regions compared to those observed in the gaseous component. The~arms, developing in the dust component spatially correlate with the gaseous arms. However, the~spiral arms outlined by the dust are narrower and demonstrate more branching compared to the gaseous spirals. A~model with twenty percent of dust admixture has a greater radial extent compared to that in the models with five and ten percent admixture of dust. We find also that the size of the particles is of great importance for the development and formation
of the nonlinear spirals in the mini-disk. The~small-sized dust particles are ``frozen'' in gas so that the dusty-gaseous mini-disk remains stable. We find that instability develops in the mini-disk if the size of the dust particles is more than 1 $\upmu$m. 

For more realistic consideration, one should take into account the size distribution of the dust particles, as~well as the processes of the formation and the destruction of the dust in the mini-disk. The~rate of the accretion of gas and dust onto the central black hole was also out of the scope of our consideration in this paper. We hope to return to these questions in future~studies.

%%%%%%%%%%%%%%%%%%%%%%%%%%%%%%%%%%%%%%%%%%
 
\vspace{6pt} 

%%%%%%%%%%%%%%%%%%%%%%%%%%%%%%%%%%%%%%%%%%

%%%%%%%%%%%%%%%%%%%%%%%%%%%%%%%%%%%%%%%%%%
\authorcontributions{Conceptualization, V.K. and R.T.; methodology, V.K. and  R.T.;
software, R.T. and  B.J.; writing—original draft preparation, R.T. and  V.K.; writing—review and
editing, R.T., V.K., and  B.J. All authors have read and agreed to the
published version of the~manuscript.}

\funding{R.T. 
thanks the Foundation for the Advancement of Theoretical Physics and Mathematics ``BASIS'' for financial support at \url{https://basis-foundation.ru}, accessed on 1 July~2022.}

\institutionalreview{Not applicable.}

\informedconsent{Not applicable.}

\dataavailability{All data used in this paper are taken from open sources and the
references are given.} 

\acknowledgments{We thank  Giovanni Carraro and  William van Altena for their careful reading of the manuscript and valuable comments. We would like to thank the anonymous reviewers for their careful reading of the article and valuable~comments.}

\conflictsofinterest{The authors declare no conflict of~interests.} 

%%%%%%%%%%%%%%%%%%%%%%%%%%%%%%%%%%%%%%%%%%

\end{paracol}

\reftitle{References}

\end{document}